\documentclass[11pt]{article}

\usepackage{amsmath}
\usepackage{amssymb}

\usepackage{graphicx}

\usepackage{cite}

\usepackage{color} 

\usepackage[labelfont=bf,labelsep=period,justification=raggedright]{caption}

\makeatletter
\renewcommand{\@biblabel}[1]{\quad#1.}
\makeatother

\date{}

\usepackage{bm}

\usepackage{color}

\newcommand{\EQ}{Eq.~}
\newcommand{\EQS}{Eqs.~}
\newcommand{\FIG}{Fig.~}
\newcommand{\FIGS}{Figs.~}
\newcommand{\CB}[1]{\textcolor{blue}{#1}}
\newcommand{\comb}[1]{#1}

\begin{document}

\begin{flushleft}
{\Large
\textbf{Structure of Cell Networks Critically Determines Oscillation
 Regularity}
}
\\
Hiroshi Kori$^{1,2,\ast}$, 
Yoji Kawamura$^{3}$, 
Naoki Masuda$^{4,2,\ast}$
\\
{\bf 1} Division of Advanced Sciences, Ochadai Academic Production, Ochanomizu University, Tokyo 112-8610, Japan
\\
{\bf 2} PRESTO, Japan Science and Technology Agency, Kawaguchi 332-0012, Japan
\\
{\bf 3} Institute for Research on Earth Evolution, Japan Agency for Marine-Earth Science and Technology, Yokohama 236-0001, Japan
\\
{\bf 4} Department of Mathematical Informatics, The
 University of Tokyo, Tokyo 113-8656, Japan
\\
$\ast$ E-mail: kori.hiroshi@ocha.ac.jp \& masuda@mist.i.u-tokyo.ac.jp
\end{flushleft}

\section*{Abstract}
%
Biological rhythms are generated by pacemaker organs,
such as the heart pacemaker organ (the sinoatrial node) and the master clock of the
circadian rhythms (the suprachiasmatic nucleus), which are
composed of a network of autonomously oscillatory cells.
Such biological rhythms have notable periodicity
despite the internal and external noise present in each cell.
Previous experimental
studies indicate that
the regularity of oscillatory dynamics is enhanced when noisy oscillators
interact and become synchronized.
This effect, called the collective enhancement of temporal precision,
has
been studied theoretically using particular assumptions.
In this study, we propose a general theoretical framework 
that enables us to understand the dependence of temporal precision 
on network parameters including size, connectivity, and coupling
intensity;
this effect has been poorly understood to date.
Our framework is based on a phase oscillator model
that is applicable to general oscillator networks with any coupling mechanism
if coupling and noise are sufficiently weak.
In particular, we can manage general directed and weighted networks.
We quantify the precision of 
the activity of a single cell and the mean activity
of an arbitrary subset of cells.
We find that, in general undirected networks, the standard
deviation of cycle-to-cycle periods scales
with the system size
$N$ as $1/\sqrt{N}$, but only up to a certain system size $N^*$
that depends on network parameters.
Enhancement
of temporal precision is ineffective when $N>N^*$.
We also reveal the advantage of long-range interactions among cells
to temporal precision.


\section*{Author Summary}
Various endogenous biological rhythms in our body such as 
heartbeats and sleep-waking cycles of about 24-hour period,
the so-called circadian rhythm, function in our body.
Unexpectedly, these rhythms maintain time regularly. For
example, the daily onset of activity in mice
has a standard deviation of a few minutes even in the absence of
environmental information. 
These biological rhythms are generated by pacemaker organs composed of
a network of autonomously oscillatory cells.
How do biological cells generate highly precise rhythms despite 
internal and external noise present in each cell?
We know, experimentally, that
an isolated cell cannot generate such precise oscillation,
but a network of coupled cells can.
Regularity in oscillations increases with the number of 
cells that constitute the network.
This effect is called the collective enhancement of temporal precision.
In this study, we present a new theory
for quantifying temporal precision in terms of
network parameters
including the number of cells, connectivity, and coupling strength.
Our main finding is that the collective enhancement is
ineffective beyond a certain cell number, and this number increases with coupling strength
among cells.
Our theory provides a useful tool for inferring the properties of cell
networks.

\section*{Introduction}
Biological rhythms such as heartbeats and sleep-waking
%
%
cycles are essential in living organisms.
%
%
Many biological rhythms are generated by pacemaker organs composed of
autonomously rhythmic cells.
For example,
the heart pacemaker (i.e., the sinoatrial node) is the source of
electric waves propagating from within
the heart, which cause the contraction of cardiac cells \cite{glass01}
The suprachiasmatic nucleus (SCN), which is a network of clock cells located in the brain,
orchestrates the circadian (i.e., approximately 24 h) activity of the entire body.
Each clock cell has a circadian rhythm in its
electric activity owing to the gene regulatory network within the cell,
and a population of clock cells synchronizes its activity through
neural interactions \cite{reppert02}.
The medullary pacemaker nucleus in electric fish is the pacemaker for the electric
discharges emitted by electric fish, which are used for object detection and
communication with other electric fish \cite{heiligenberg81}.

Cell dynamics involve fluctuations resulting from
various types of internal and external noise. However,
oscillations in pacemaker organs
such as the sinoatrial node in the heart, the SCN,
and the medullary pacemaker nucleus in electric fish
are highly precise.
For example, the daily onset of activity in certain mammals and birds
has a standard deviation (SD) of a few minutes even in the absence of environmental information
\cite{enright80}. In addition, the electric organ discharge 
pattern in certain electric fish has a standard deviation of as little as $0.02\%$
of the average period \cite{moortgat00}. 

Experiments by Clay and DeHaan provided an important clue for
understanding the mechanisms underlying precise oscillations was
provided by \cite{clay79}.
They prepared clusters of cultivated cardiac
cells, ranging in size from $1$ to $\sim$$100$, and observed the beatings of
individual cells. They found that the SD of
inter-beat intervals decreases with the number of component cells
in the cluster ($N$) roughly as ${\rm SD} \propto 1/\sqrt{N}$.
Therefore, precision in individual cell oscillations is enhanced as the
number of cells increases.
Note that this scaling,
which is reminiscent of the central limit theorem,
is not at all trivial.
This is because
oscillators are synchronized and thus strongly correlated, while
the central limit theorem is applicable to an ensemble of independent 
elements.

The decrease in SD as $N$ increases,
 the so-called collective enhancement of
temporal precision, has attracted considerable attention
\cite{enright80, winfree01, clay79, moortgat00, herzog04, kojima06,
sherman88, moortgat00b,garcia04pnas, vasalou09, rappel96,needleman01,ly10,tabareau10}.
There is a large body of experimental
\cite{clay79, moortgat00, herzog04, kojima06}, numerical 
\cite{sherman88, moortgat00b,garcia04pnas, vasalou09},
and analytical
\cite{rappel96,needleman01,ly10,tabareau10} studies.
Theoretically, it has been shown that
the average activity of all oscillators 
on the all-to-all network (i.e., the complete graph)
obeys ${\rm SD} \propto 1/\sqrt{N}$ \cite{rappel96, needleman01}.
However, most analytical studies are based on rather strong assumptions about
coupling topology (e.g., all-to-all) or coupling mechanism (e.g.,
gap-junction type). 
Moreover, little is known about temporal precision
in single cell activity or ensemble activity for
a subset of cells in an entire network.
Note that in the experiments by Clay and DeHaan, 
the behavior ${\rm SD} \propto 1/\sqrt{N}$ was found for
single cells and not for the entire network \cite{clay79}.



In this paper, we propose 
a general theoretical framework that enables us to understand
the dependence of temporal precision 
on network parameters, including size, connectivity, and coupling intensity.
Our framework, based on a phase oscillator model,
allows us to handle directed
and weighted networks, various coupling mechanisms, and
temporal precision in the activity of single
cells and arbitrary subsets of cells.

We begin by describing the numerical results for two biological
pacemaker models: network of the FitzHugh-Nagumo oscillators and 
that of circadian oscillators.
These models have distinct oscillation and coupling mechanisms.
For different networks including all-to-all coupling,
lattices with nearest-neighbor coupling,
and the random graph,
we observe that there is a common dependence of
temporal precision on network size $N$.
The SD of cycle-to-cycle periods decreases 
as $1/\sqrt{N}$ in small networks, but approaches an
asymptotic value
%
%
as $N$ increases. That is, there is a crossover.
Then, we develop a theory for obtaining an explicit expression for the
SD of the cycle-to-cycle period.
In particular, we find the condition for the behavior ${\rm SD} \propto
1/\sqrt{N}$
and the dependence of the crossover point $N^*$ on network parameters.
We also demonstrate the advantage of long-range interactions among cells
to temporal precision.
Finally, we discuss the implications of our theory.

\section*{Results}

\subsection*{Numerical results} \label{sec:numerical}
First, we present the numerical results
for two mathematical models describing biological
oscillations (see {\bf Methods} for the details of the models).
We used
FHN oscillators with gap-junction coupling as a model of oscillatory
cardiac or neural cells. 
We also employed a previously proposed model for the SCN (i.e., a population of
circadian clock cells) \cite{locke08},
which is referred to as the SCN model.


\subsubsection*{Waveforms and oscillation periods are regularized when 
   oscillators are coupled}
\begin{figure}[!ht]
\begin{center}
\includegraphics[width=6in]{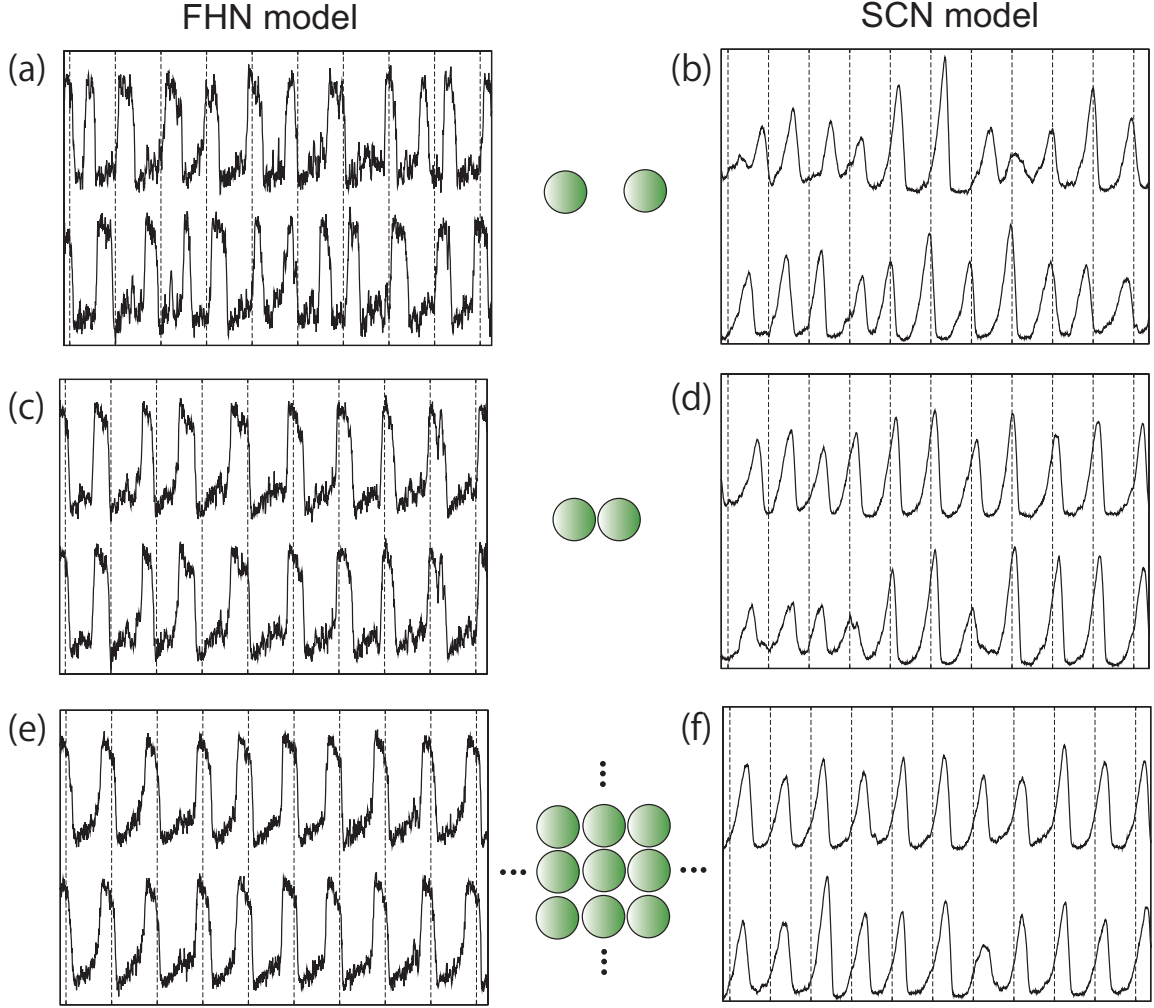}
\end{center}
\caption{
{\bf Waveforms obtained from biological oscillator models.} We
present the time series of $x_i(t)$
of (a,b) two isolated cells ($\kappa=0, N=2$),
(c,d) two coupled cells ($\kappa>0, N=2$), and (e,f) two
cells in 100 coupled cells ($\kappa>0, N=100$) in
(a,c,e) the FHN model and (b,d,f) the SCN model.
In (e), we employ
the one-dimensional lattice with an open boundary condition
and show the waveforms of two neighboring cells.
In (f), we employ the all-to-all network
($A_{ij}=1/N$ for $1 \le i,j \le N$).
We set $\rho=0.1$ in all panels and (a,b) $\kappa=0$, (c,e)
 $\kappa=2$, and (d,f) $\kappa=1$.
}
\label{fig:waveforms}
\end{figure}

Figures~\ref{fig:waveforms}(a,c,e) and (b,d,f) present the waveforms
obtained from the FHN and SCN models of different network sizes, respectively.
The average cycle-to-cycle periods are depicted by 
dotted lines in each panel to illuminate the variations in
cycle-to-cycle periods.
The properties of each constituent cell were kept constant,
while the connectivity between the cells is different.
Typical waveforms of uncoupled cells ($N=2$)
are shown in \FIGS\ref{fig:waveforms}(a,b).
When the cells are coupled sufficiently strongly, the
system synchronizes stably (\FIGS\ref{fig:waveforms}(c,d)).
Figures~\ref{fig:waveforms}(c,d) indicate that waveforms
in the presence of coupling
are regularized as compared to the waveforms of isolated cells [\FIGS\ref{fig:waveforms}(a,b)].
In particular,
the variation in the cycle-to-cycle period decreases.
When 100 oscillators are coupled
[\FIGS\ref{fig:waveforms}(e,f)], the variation appears to be even smaller.
When cells are coupled, individual cell oscillations are
not only synchronized but also regularized, and the
oscillation appears to be more regular
for a larger system size.


\subsubsection*{There is a limit to the enhancement of temporal precision}
\begin{figure}[!ht]
\begin{center}
\includegraphics[width=4in]{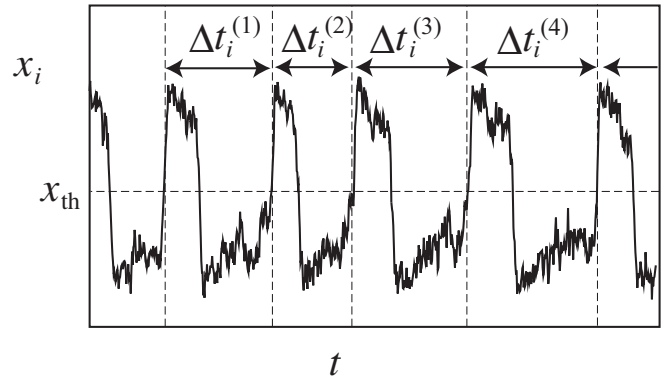}
\end{center}
\caption{
{\bf Schematic illustration of the concept of cycle-to-cycle
 period.}
}
\label{fig:time_series_dt}
\end{figure}

To quantify the dependence of temporal precision on
network parameters, we measured the coefficient of variation (CV), which
is the SD of the cycle-to-cycle period divided by the
mean period.
A cycle-to-cycle period is defined by an interval $\Delta t$
between two successive passages of an observed variable ($x_i$)
across a specified threshold value $x_{\rm th}$ (\FIG\ref{fig:time_series_dt}).
We set $x_{\rm th}=0.4$ and $2.0$ for the FHN and SCN models,
respectively.
We discard $\Delta t$ that is much smaller than
a typical oscillation period to exclude
noise-driven rapid threshold crossing.
The CV is defined as
\begin{equation}
 {\rm CV} = \frac{\rm SD}{\rm \tau},
\end{equation}
where $\tau$ and SD are the mean and the SD of a series of
$\Delta t$, respectively.

%

Here we investigate the FHN model
on networks of different types and different sizes.
We assume that the system is composed of identical cells subjected to weak noise.
Figure \ref{fig:fhn}(a) shows the CV of individual cell oscillations in the
FHN model on the all-to-all network of size $N$. 
The results for different coupling strength 
values, $\kappa$, are plotted using different symbols.
We find that 
\begin{itemize}
 \item[(i)] CV is proportional to $1/\sqrt{N}$ for small $N$ values for
	    each $\kappa$
 \item[(ii)] CV approaches a constant
value for large $N$ values for each $\kappa$; i.e.,
there is a crossover
\item[(iii)] the crossover point $N^*$ increases with $\kappa$.
\end{itemize}

We observe similar behavior
for the square lattice and the random graph, as shown in
\FIGS\ref{fig:fhn}(b) and (c), respectively.

\subsubsection*{Temporal precision increases with $N$, while the level
   of synchrony remains constant}

A natural question is whether the enhanced synchronization induces
the collective enhancement of temporal
precision.
%
%
To examine this possibility, 
we measured the distance $\delta$ between the actual state
and the in-phase state (see {\bf Methods} for the definition of $\delta$)
for the all-to-all network.
As shown in \FIG\ref{fig:fhn}(d),
the level of synchrony is independent of $N$ for each $\kappa$ value.
We also confirmed that, in the FHN model on a square lattice,
$\delta$ even increases with $N$ although the CV decreases (results not shown).
Thus, the enhancement of temporal precision 
by an increase in $N$ is not attributed to the improvement in
synchronization.

\subsubsection*{CV for ensemble activity has a larger crossover point}
In nature,
rhythmic output from a pacemaker organ is usually
generated by an ensemble of multiple cells.
For example, rhythmic electroactivity propagating within the heart
is thought to originate from cells on the surface of the
sinoatrial node.
The SCN consists of various neural populations, 
and each population forms a particular pattern of efferent projections
to other parts of the brain \cite{abrahamson01}.
This anatomical fact suggests that the SCN's output is generated by
a combination of a subset of neurons
rather than by the uniform average of the entire organ.

Therefore,
we investigated the CV of the ensemble activity of a subset of cells
on the all-to-all network.
The ensemble activity is defined by the average waveform of $M$
($1\le M\le N$) cells:
\begin{equation}
X(t) = \frac{1}{M} \sum_{i=1}^{M} x_i(t),
\end{equation}
where 
the measured ensemble is assumed to consist of oscillators $x_1$,
$\ldots$, $x_M$.
The cycle-to-cycle period and the CV for the ensemble activity are
defined similarly to
the case of single cell activity (\FIG\ref{fig:time_series_dt}).
In \FIG\ref{fig:fhn_multi}, we
present the CV measured for the average waveform with different values of $M$
in the FHN model
on the all-to-all network. 
For $M$ smaller than $N$, 
properties (i)--(iii) listed above are preserved.
In addition, we find that
\begin{itemize}
 \item[(iv)] the crossover point $N^*$ increases with ensemble size $M$
 \item[(v)] for $M=N$, the CV is proportional to $1/\sqrt{N}$ for
any $N$; i.e., there is no crossover.
\end{itemize}
We also confirmed that the same properties hold true
for the FHN model on two--dimensional lattices and for the
SCN model
on the all-to-all network and the two--dimensional lattice.
%

\subsubsection*{Results are qualitatively the same under strong noise and heterogeneity}
So far, we have assumed an ideal case: identical oscillators
and weak noise.
To simulate more realistic situations, we now consider networks composed
of heterogeneous cells subjected to relatively strong noise.
As examples, we measure the CV for the 
FHN model on the square lattice
and for the SCN model on the all-to-all network (Fig. \ref{fig:hetero}).
In the FHN model, we made
one of the parameter values heterogeneous
in order to obtain the distribution of natural periods of cells as $\tau_i \approx
133 \pm 3$ (mean $\pm$ SD). In the SCN model,
the time scales of the cells were
made heterogeneous such that
$\tau_i \approx 23.4 \pm 1.2$. The latter situation is
consistent with the experimental observation by Honma et
al. \cite{honma98}.
In all cases, we apply
sufficiently strong coupling to ensure that the
oscillators are well synchronized.
Under this condition, as seen in \FIG\ref{fig:hetero},
all properties (i)--(v) hold true.

%

\subsection*{Theory}


We found, numerically, that properties (i)--(v) hold true in various
situations. 
In the following,
we develop a theory for relating temporal precision to network parameters
by assuming weak coupling and weak noise. Under this assumption,
a large class of oscillator systems including the 
models considered above
are reduced to the phase model (see {\bf Methods} and
References \cite{winfree67,kuramoto84})
%
%
given by
\begin{equation}
 \dot \phi_i(t) = \omega_i +
 \kappa \sum_{j=1}^N A_{ij} f(\phi_j-\phi_i)+ \sqrt{D} \xi_i(t),
 \label{model}
\end{equation}
where $\phi_i$ and $\omega_i$ $(1\le i\le N)$ are the
phase and intrinsic frequency of the $i$th oscillator, respectively;
$A=(A_{ij})$
is the weighted adjacency matrix with its element $A_{ij}$ equal to
the intensity of the coupling from the $j$th to $i$th oscillators; $\kappa$ is the
overall coupling intensity; $f(\cdot)$ is a $2\pi$--periodic function;
$\xi_i(t)$ is independent white Gaussian noise with ${\rm
E}[\xi_i(t)]=0$ and ${\rm E}[\xi_i(t) \xi_j(t')] =
\delta_{ij}\delta(t-t')$, where ${\rm E}$ represents the expectation;
and $D$ is the strength of the noise.
The adjacency matrix $A$
is allowed to be asymmetric, weighted, and to possess negative 
components. Extension of the following results in the case of
$i,j$-dependent coupling function $f_{ij}(\cdot)$
and $i$-dependent noise strength $D_i$ is straightforward. For
clarity of the presentation,
we focus on Eq.~\eqref{model}.
We assume that all the oscillators are synchronized in frequency;
i.e., all the oscillators have the actual frequency $\Omega$ owing to the
effect of coupling.
Synchronization usually occurs when coupling is sufficiently strong
compared to noise and heterogeneity in $\omega_i$.

\begin{figure}[!ht]
 \begin{center}
  \includegraphics[width=2.5in]{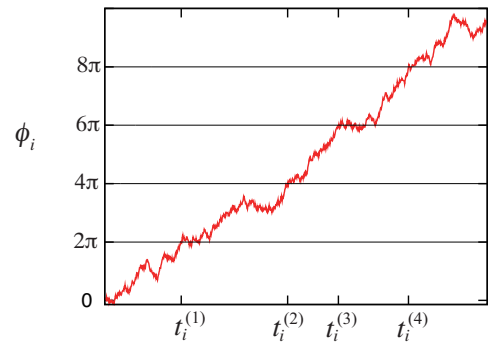}
 \end{center}
 \caption{
{\bf First passage time for a phase oscillator.}
}
\label{fig:first_passage}
\end{figure}

One oscillation cycle corresponds to
an increase in the phase by $2\pi$.
More precisely, the $k$th cycle-to-cycle period of the $i$th
oscillator is defined by $\Delta t_i^{(k)} = t_i^{(k)}-t_{i}^{(k-1)}$, where 
$t_i^{(k)}$ is the first passage time for $\phi_i(t)$ to
exceed $2k\pi$ (Fig.~\ref{fig:first_passage}).
Because we assumed that all the oscillators are
synchronized to $\Omega$, 
the expected value of $\Delta t_i^{(k)}$ ($\tau$)
is independent of $i$ and is given as
\begin{align}
 \tau \equiv {\rm E}[\Delta t_i^{(k)}] = \frac{2\pi}{\Omega},
\end{align}
where the statistical averages are taken over different $k$ values.
The temporal precision of the $i$th oscillator is characterized by
\begin{align}
 {\rm SD}_i &\equiv {\rm std}[\Delta t_i] = \sqrt{{\rm
var}[\Delta t_i]} = \sqrt{{\rm E}[ (\Delta t_i^{(k)} - \tau)^2 ]}.
\end{align}
The CV for the $i$th oscillator is equal to
\begin{equation}
 {\rm CV}_i \equiv \frac{{\rm SD}_i}{\tau}.
  \label{CV_i_def}
\end{equation}
%

\begin{figure}[!ht]
 \begin{center}
  \includegraphics[width=6in]{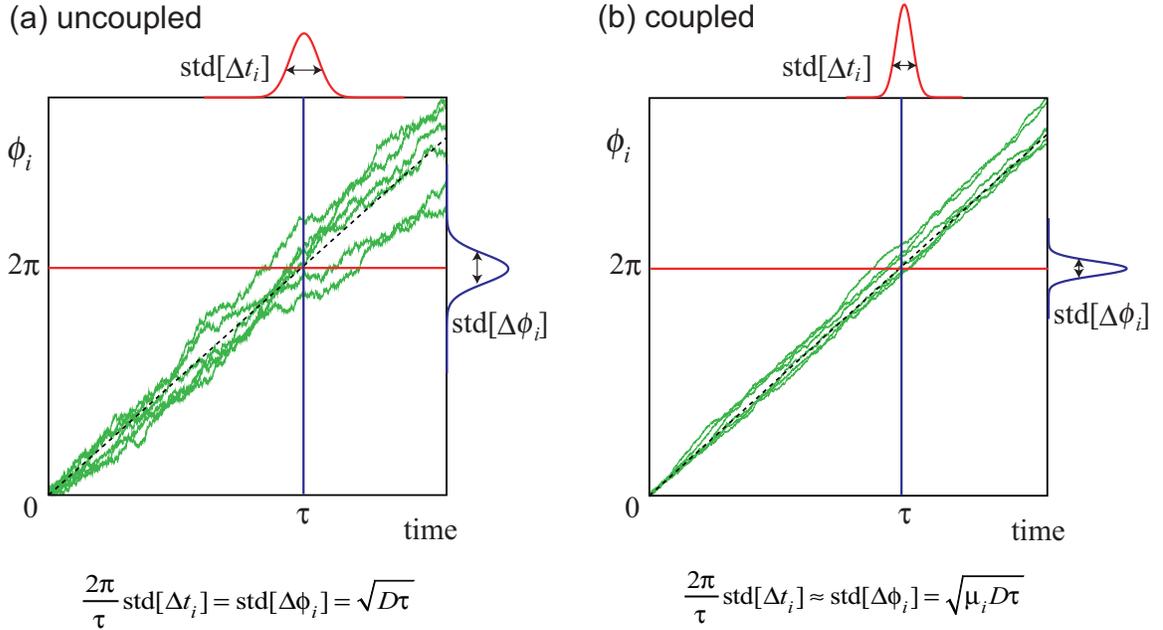}
 \end{center}
 \caption{ 
 {\bf Schematic illustration of our approximation to the 
cycle-to-cycle variation.}
Green trajectories represent different realizations of
the phase $\phi_i(t)$ of a single oscillator
in (a) uncoupled and (b) coupled cases, where we set $\phi_i(0)=0$.
The red curves on the top of each panel represent distribution function
$P_t(t)$
obtained from the first passage time of $\phi_i(t)=2\pi$
in different realizations.
Our concern is its standard deviation, ${\rm SD}_i = {\rm std}[\Delta t_i]$.
The blue curves on the right of each panel represent
distribution function $P_\phi(\phi_i)$ 
obtained from different realizations of $\phi_i(\tau)$, where $\tau$ is the mean
period, and its standard deviation is denoted by ${\rm std}[\Delta \phi_i]$.
We approximate ${\rm std}[\Delta t_i]$ using ${\rm std}[\Delta \phi_i]$.
Suppose that $P_t( t) dt = P_\phi( \phi_i) d \phi_i$.
On average, the phase crosses $2\pi$
with slope $2\pi/\tau$, i.e.,
$d\phi_i/dt = 2\pi/\tau$. We thus obtain \EQ\eqref{heuristic}.
For uncoupled oscillators ($\kappa=0$), our model corresponds to the
Wiener process with a constant drift. In this case,
\EQ\eqref{heuristic} is exact, and we obtain $(2\pi/\tau) {\rm
 std}[\Delta t_i] = {\rm std}[\Delta \phi] = \sqrt{D\tau}$ \cite{gerstner02}.
We also know that \EQ\eqref{heuristic} is asymptotically exact in the one-dimensional Ornstein-Uhlenbeck
process for weak noise \cite{gerstner02}. 
For coupled oscillators $(\kappa>0)$, however, 
our model (\ref{model}) is a multivariate Ornstein-Uhlenbeck process
when linearized.
Even in this case, as is numerically confirmed in the
examples shown in \FIGS\ref{fig:n2}--\ref{fig:sw}, \EQ\eqref{heuristic}
provides a suitable approximation.
} \label{fig:heuristic}
\end{figure}

To obtain the dependence of ${\rm SD}_i$ on network parameters,
we employ an approximation given by
\begin{equation}
\frac{2\pi}{\tau} {\rm std}[\Delta t_i]\approx {\rm std}[\Delta
  \phi_i],
  \label{heuristic}
\end{equation}
where $\Delta \phi_i \equiv \phi_i(t+\tau) - \phi_i(t) - 2\pi$
(Fig. \ref{fig:heuristic}).
%
For an isolated oscillator 
obeying $\dot \phi_i = \omega_i + \sqrt{D}\xi_i(t)$, one immediately
finds that
${\rm var}[\Delta \phi_i]= D \tau_i$, where $\tau_i=2\pi/\omega_i$.
When oscillators are coupled and synchronized with frequency $\Omega$,
we write
\begin{equation}
 {\rm var}[\Delta \phi_i]= \mu_i D \tau.
\end{equation}
We refer to $\mu_i$ as the scaling factor
of the $i$th
oscillator (\FIG\ref{fig:heuristic}).

To obtain an expression for $\mu_i$,
we assume that noise is sufficiently weak
and linearize \EQ\eqref{model} around the synchronized state.
The synchronized solution $\phi_i^{\rm s}(t)$ $(1\le i \le N)$ is represented as
\begin{equation}
\phi_i^{\rm s}(t) = \Omega t + \psi_i, 
\label{sync}
\end{equation}
where $\Omega$ and $\psi_i$ are the constants derived by setting $\dot
\phi_i = \Omega$ and $D=0$ in \EQ\eqref{model}; i.e.,
\begin{equation}
\Omega = \omega_i + \kappa \sum_{j=1}^N A_{ij} f(\psi_j - \psi_i).
\end{equation}
By introducing a small deviation
\begin{equation}
 \theta_i(t)=\phi_i(t)- \phi_i^{\rm s}(t),
  \label{deviation}
\end{equation}
we obtain
\begin{equation}
 \dot \theta_i(t)= \kappa \sum_{j=1}^N w_{ij}(\theta_j-\theta_i)+\sqrt{D} \xi_i(t),
\label{linear2}
\end{equation}
where $w_{ij} = A_{ij} f^{\prime}(\psi_j-\psi_i)$ is the effective
coupling weight. For convenience, we rewrite \EQ\eqref{linear2} as
\begin{equation}
 \dot \theta_i(t)= - \kappa \sum_{j=1}^N L_{ij}\theta_j +\sqrt{D} \xi_i(t),
\label{linear}
\end{equation}
where $L=(L_{ij})$ is the Jacobian matrix with its element $L_{ij}$
given by
\begin{equation}
 L_{ij} = \left\{
	  \begin{array}{cl}
	   - w_{ij} & \mbox{for $i \neq j$},\\
	   \sum_{i' \neq i} w_{ii'} & \mbox{for $i = j$}.\\
	  \end{array}
\right.
\end{equation}
Note that $L$ has a zero eigenvalue with the corresponding right
eigenvector $\bm u^{(1)}=(1,
\ldots, 1)^\top/\sqrt{N}$.
Furthermore, because of the assumption of the stability of the
synchronized state, the real parts of the other $N-1$ eigenvalues of $L$
are positive, \textit{i.e.}, $0\equiv \lambda_1 < {\rm Re} \lambda_2 \le
\ldots \le {\rm Re} \lambda_N$.  The assumption of the stability holds
true when $w_{ij}\ge 0$ ($1\le i, j\le N$) and the network described by
the adjacency matrix $(w_{ij})$ is strongly connected
\cite{Ermentrout92,Agaev00,Arenas08}.  For more general cases, the
stability condition is nontrivial.

For in-phase synchrony (i.e.,
$\psi_i=0$ for $1\le i \le N$ in \EQ\eqref{sync}), which occurs when the
heterogeneity in the network and in individual
oscillators is sufficiently
small and/or the coupling is sufficiently strong, 
we obtain $w_{ij} \propto A_{ij}$ for $1\le i,j \le N$.
In this case, $L$ is the network Laplacian generalized for
a directed and weighted network \cite{newman10}, given by
\begin{equation}
 L_{ij} \propto \left\{
	  \begin{array}{cl}
	   - A_{ij} & \mbox{for $i \neq j$},\\
	   \sum_{i' \neq i} A_{ii'} & \mbox{for $i = j$}.\\
	  \end{array}
\right.
\end{equation}
Note that $L$ is symmetric when the adjacency matrix $A$ is
symmetric.  

As shown in {\bf Methods}, for any diagonalizable matrix $L$, we obtain
$\mu_{i} = C_{ii}$, where
\begin{align}
   C_{ij}  
&\equiv \frac{{\rm E}[(\theta_i(t+\tau)-\theta_i(t)) (\theta_j(t+\tau)-\theta_j(t))]}{D\tau} \notag\\
&=
\frac{V_{11}}{N} +  \sum_{m,n\; (m+n > 2)}^N
\frac{2-e^{-\kappa \lambda_m \tau}-e^{-\kappa \lambda_n
   \tau}}{\kappa(\lambda_m+\lambda_n)\tau}
 V_{mn} u^{(m)}_i u^{(n)}_j.
 \label{C_ij}
\end{align}
Here $\bm u^{(n)}=(u^{(n)}_i)$ and $\bm v^{(n)}$ are, respectively,
the right and left eigenvectors of $L$
that satisfy the orthogonality and normalization
conditions; i.e., $L\bm u^{(n)}=\lambda_n \bm
u^{(n)}$, $\bm v^{(n)} L = \lambda_n \bm v^{(n)}$, and
$\bm v^{(m)} \bm u^{(n)}=\delta_{mn}$;
and $V_{mn} = \bm v^{(m)} \cdot \bm v^{(n)}$.

For symmetric $L$, which is the case for in-phase synchrony on
undirected networks, \EQ\eqref{C_ij} becomes much simpler.
Because all the eigenvalues are real,
$\bm u^{(n)}=\bm v^{(n)\top}$, $V_{mn} = \bm u^{(m)} \cdot \bm
u^{(n)}=\delta_{mn}$ for $1 \le m,n \le N$, and 
$\sum_{i=1}^N u^{(n)}_i \propto \bm u^{(1)} \cdot \bm u^{(n)}=0$ 
for $n \ge 2$, we obtain
\begin{align}
   \mu_{i} = 
 \frac{1}{N} + \sum_{n=2}^N 
  \frac{1-e^{-\kappa \lambda_n \tau}}{\kappa \lambda_n \tau}
 u_i^{(n)} u_i^{(n)}.
 \label{undirected}
\end{align}
Moreover, because of the normalization condition,
$\sum_{i=1}^{N} u_i^{(n)} u_i^{(n)}=1$,
the mean of $\mu_{i}$ over the entire network, $\langle \mu \rangle
=\sum_{i=1}^N \mu_{i}/N$, is independent of the eigenvectors
and is given by
\begin{align}
   \langle \mu \rangle= 
 \frac{1}{N} + \frac{1}{N} \sum_{n=2}^N 
  \frac{1-e^{-\kappa \lambda_n \tau}}{\kappa \lambda_n \tau}.
 \label{undirected2}
\end{align}

\subsubsection*{Crossover point $N^*$ increases with coupling strength $\kappa$}

If the second
term of \EQ\eqref{undirected2}
is negligible compared to the first term, we obtain $\langle \mu \rangle
\approx 1/N$; i.e., the SD decreases proportionally to $1/\sqrt{N}$.
However, as $N$ increases, the second term becomes
comparable at certain $N^*$ and even dominant for $N \gg N^*$.
If the eigenvalue spectrum converges to a certain density function
$q(\lambda)$ as $N \to \infty$, we obtain
\begin{equation}
 \langle \mu \rangle \to \mu_\infty \equiv \int_{0}^{\infty} q(\lambda)
  \frac{1-e^{-\kappa \lambda \tau}}{\kappa \lambda \tau} d \lambda \quad
  (N \to \infty) .
\end{equation}
We later demonstrate the convergence for the
all-to-all and ring networks. Spectra of finite dimensional
lattices \cite{mohar91}, 
uncorrelated random graphs with arbitrary degree
distributions \cite{samukhin08}, and the small-world
network with a fixed expected degree \cite{monasson99} also converge.
By equating the first and second terms in
\EQ\eqref{undirected2}, we estimate the crossover point 
as $N^* \sim 1/\mu_\infty$.
Since $\mu_\infty$ monotonically decreases with increasing
$\kappa$, $N^*$ increases with $\kappa$.

Furthermore, if the second smallest eigenvalue $\lambda_{2}$ 
is nonvanishing in the limit $N \to \infty$ (which is the case, for
example, in the all-to-all
network and various random networks including small-world
networks \cite{monasson99, samukhin08}) and $\kappa$ is so large that
$e^{-\kappa \lambda_{2} \tau} \ll 1$, we obtain $\mu_\infty \propto 1/\kappa$.
Then, the crossover point scales as 
\begin{equation}
 N^* \propto \kappa. 
  \label{N_propto_kappa}
\end{equation}

\subsubsection*{Crossover point $N^*$ is proportional to the size of a
   measured ensemble}
By assuming in-phase synchrony,
we calculate the scaling factor
of the noise reduction for the ensemble activity of
an arbitrary set of oscillators.
%
%
We rearrange the oscillator indices and write
the ensemble activity as
\begin{equation}
 X(t) = \sum_{i=1}^M \zeta_i x_i(t),
  \label{ensemble_activity}
\end{equation}
where $\zeta_i \ge 0$ is
an arbitrary constant 
with the normalization
condition $\sum_{i=1}^M \zeta_i = 1$.
When the deviation $\theta_i$ from in-phase synchrony (i.e., $\psi_i=0$
for $1\le i \le N$ in \EQ\eqref{sync})
is small for each oscillator, the phase of $X(t)$ is approximated by
\begin{equation}
 \Phi (t) = \sum_{i=1}^M \zeta_i \phi_i(t) = \Omega t+ \sum_{i=1}^M \zeta_i \theta_i(t).
  \label{mean_phase}
\end{equation}
Then, similar to the case of individual cell oscillations,
we define the scaling factor $\mu_\Phi$ for the ensemble activity as
\begin{equation}
 {\rm var}[\Delta \Phi] = \mu_\Phi D \tau,
\end{equation}
where $\Delta \Phi = \Phi(t+\tau) - \Phi(t) - 2\pi$.
We then obtain
\begin{equation}
 \mu_\Phi
 = \frac{{\rm var}[\Delta \Phi]}{D \tau} 
 = \sum_{i,j=1}^M \zeta_i \zeta_j \frac{{\rm E}[
(\theta_i(t+\tau)-\theta_i(t)) (\theta_j(t+\tau)-\theta_j(t)) ]}{D \tau}
 = \sum_{i,j=1}^M  \zeta_i \zeta_j C_{ij}.
  \label{mu_Phi}
\end{equation}
Henceforth, we assume $\zeta_i=1/M$ for $1\le i \le M$,
as is the case
in \FIGS\ref{fig:fhn_multi} and \ref{fig:hetero}(b).

There are notable properties for symmetric $L$ (see {\bf Methods}).
When $M=N$ (i.e., $X(t)$ is the mean activity of the entire network), we obtain
\begin{align}
   \mu_{\Phi} = \frac{1}{N},
 \label{mu_Phi_N}
\end{align}
that is, there is no crossover.  For $M < N$, $\mu_{\Phi}$ generally
depends on the choice of $M$ oscillators. However,
if we randomly choose $M$ 
oscillators out of $N$ oscillators, where $1\ll M \ll N$, we estimate
\begin{align}
   \mu_{\Phi} \approx
 \frac{1}{N} + \frac{1}{MN} \sum_{n=2}^N
 \frac{1-e^{-\kappa \lambda_n \tau}}{\kappa \lambda_n \tau}
 \approx \frac{1}{N} + \frac{\mu_{\infty}}{M}.
 \label{undirected_ensemble2}
\end{align}
In this case, the lower bound of the SD
is inversely proportional to $\sqrt{M}$ and the crossover point increases as
\begin{equation}
 N^* \propto M.
\end{equation}
As shown later, this estimation is asymptotically exact for 
the all-to-all network.


\subsubsection*{Behavior $1/\sqrt{N}$ can be
   violated even for small $N$ values when $L$ is asymmetric}
The behavior ${\rm CV} \propto 1/\sqrt{N}$ is obtained for $N<N^*$
when the Jacobian $L$
is symmetric, which is the case when a network is undirected
and the oscillators are synchronized in phase.
We refer to this situation as ``democratic'' because symmetric $L$ implies 
that the action and reaction between any two nodes are balanced.

For asymmetric $L$, \EQ\eqref{C_ij} implies that
the SD at small $N$ values decreases as $\sqrt{V_{11}/N}$ instead of $1/\sqrt{N}$. 
In \cite{mkk10},
we analyzed the long-time diffusion property of \EQ\eqref{model}
to obtain 
$\sigma^2 \equiv \lim_{\Delta t \to \infty} {\rm var}[\theta_i(t
+ \Delta t) - \theta_i(t)]/(D \Delta t) =V_{11}/N$ through a different technique.
This previous result is consistent
with that obtained in the present paper
because 
$\sigma^2$ corresponds to phase diffusion after infinitely many cycles,
and the second term on the right-hand side
of \EQ\eqref{C_ij} vanishes with this limit.
Furthermore,
we showed in \cite{mkk10}
that $\sqrt{V_{11}/N}$ is
larger than or equal to $1/\sqrt{N}$ for asymmetric $L$.
For example, in directed scale-free networks,
which is a strongly heterogeneous network, we obtained
$\sqrt{V_{11}/N} \propto N^{-\beta}$ with $0 \le \beta \le
1/2$; the effect of collective enhancement is significantly weaker.
%
%
Moreover, the scaling $\sqrt{V_{11}/N}=N^{-1/2}$ can be violated 
even when a network is undirected. 
This is the case when the 
synchronized state is not in-phase but accompanies
a wave pattern. Wave patterns arise
when the network is spatially extended
(such as Euclidian lattices) and the natural frequency is sufficiently heterogeneous
\cite{kuramoto84,blasius05}. 
In this case, $V_{11}/N$ decreases with $N$ for small $N$ values but approaches
a constant value for large $N$ values.
Thus, strongly asymmetric connectivity and/or
strong heterogeneity in the oscillator's properties
can hamper the collective enhancement of temporal
precision.

\subsection*{Examples and numerical verification}

To demonstrate and numerically confirm our
analytical results, we investigate the phase model (\EQ\eqref{model})
on several networks. In numerical simulations, we set
$\omega_i=1$,
$f(\phi)=\sin \phi$, and $\sqrt{D}=0.01$ in \EQ\eqref{model}.
In the example networks, all the
oscillators synchronize in phase in the absence of noise. Thus, $w_{ij}
= A_{ij}$ and $\Omega=\omega$ for any coupling strength $\kappa$ and any
$N$.
Note that the dependence of the CV on $\kappa$ and $N$ is only through
the SD because $\tau=2\pi/\omega$ is constant. In the following, we
show the values of the normalized CV, that is 
actual CV values divided by the CV of isolated
oscillators, shown as $\sqrt{D \tau}/2\pi$.
Our theory predicts that ${\rm CV}_i \approx \sqrt{\mu_{i}}$ and
${\rm CV}_\Phi \approx \sqrt{\mu_\Phi}$.

\begin{figure}[!ht]
 \begin{center}
   \includegraphics[width=5in]{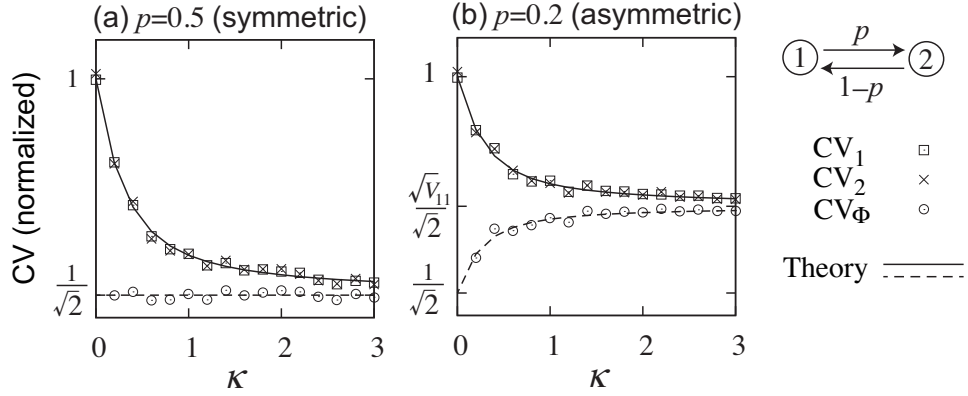}
 \end{center}
 \caption{{\bf Normalized CV versus coupling strength in asymmetrically
 coupled phase oscillators ($N=2$).}
Presented is the normalized CV, i.e., ${\rm CV}_{i}/(\sqrt{
D \tau}/2\pi)$ $(i=1,2)$ and ${\rm CV}_{\Phi}/(\sqrt{
D \tau}/2\pi)$ for two coupled phase oscillators with (a) $p=0.5$ and
(b) $p=0.2$. Numerical results are shown by symbols.
The solid and dotted lines represent the analytical results given by
\EQS\eqref{mu_ii_n2} and \eqref{mu_Phi_n2}, respectively. Note that
$V_{11}=2p^2+2(1-p)^2=1$ for $p=0.5$ and $V_{11}=1.36$ for $p=0.2$.}  \label{fig:n2}
\end{figure}

{\em Two asymmetrically coupled elements}.
The first example is two asymmetrically coupled elements ($N=2$):
$w_{12}=p$ and $w_{21}=1-p$ (\FIG\ref{fig:n2}). 
In this case, we have $\lambda_1=0, \lambda_2=-1,
\bm u^{(1)} = \frac{1}{\sqrt{2}}(1 \; 1)^\top, \bm v^{(1)} = \sqrt{2}(p \; 1-p),
\bm u^{(2)} = \sqrt{2}(1-p \; -p)^\top$, and $\bm v^{(2)} = \frac{1}{\sqrt{2}}(1\; -1)$.
By substituting
them in \EQS\eqref{C_ij} and \eqref{mu_Phi} for $M=2$ and setting
$\zeta_1=\zeta_2=1/2$, we obtain
\begin{align}
   \mu_{1}  &= \mu_{2}=
 \frac{V_{11}}{2} + \left(1-\frac{V_{11}}{2}\right)\frac{1-e^{-\kappa
 \tau}}{\kappa \tau}, \label{mu_ii_n2} \\
   \mu_{\Phi}  &= \frac{V_{11}}{2} + \left(\frac{1}{2}-\frac{V_{11}}{2}\right)\frac{1-e^{-\kappa
 \tau}}{\kappa \tau}, \label{mu_Phi_n2} 
\end{align}
where $V_{11}=2p^2+2(1-p)^2$. 
For any $\kappa$ and $\tau$ values, the best
precision is obtained in the symmetric case ($p=0.5$).
Figure~\ref{fig:n2} suggests that
the analytical and numerical results are in strong agreement.

\begin{figure}[!ht]
\begin{center}
\includegraphics[width=6in]{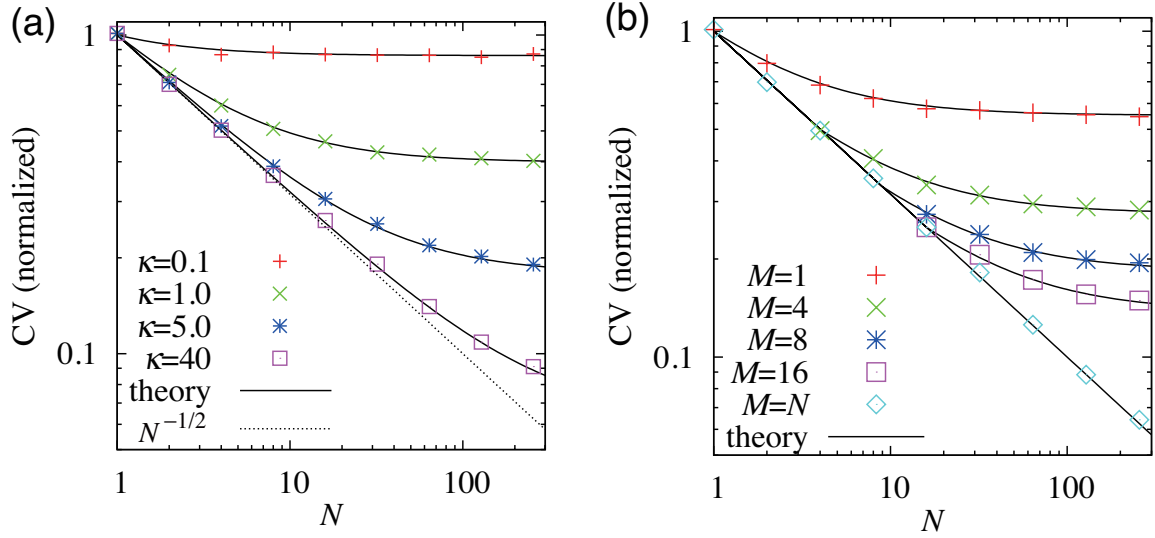}
\end{center}
 \caption{{\bf Normalized CV in phase oscillators on the all-to-all network.}
(a) CV for individual cells ($M=1$)
for various $\kappa$ and $N$ values.
(b) CV for ensemble activity for various $M$ values with $\kappa=0.5$.
Symbols represent numerical data. Solid lines represent
(a) $\sqrt{\mu_i}$ given by \EQ\eqref{mu_i_global} and
(b) $\sqrt{\mu_\Phi}$ given by \EQ\eqref{mu_phi_global}.}
 \label{fig:kuramoto_global}
\end{figure}

{\em All-to-all coupling}.
The second example is all-to-all coupling; i.e.,
$w_{ij}=1/N$ for $1 \le i,j \le N$.
The eigenvalues are given by
$\lambda_n=1$ ($2\le n\le N$). Because all the nodes
are equivalent (i.e., permutation symmetry),
we obtain $\mu_{i}= \langle \mu \rangle$. Then, from
\EQ\eqref{undirected2}, it follows that
\begin{align}
   \mu_{i}=
 \frac{1}{N} + \left(1-\frac{1}{N}\right)\frac{1-e^{-\kappa\tau}}{\kappa\tau}. 
 \label{mu_i_global}
\end{align}
We also obtain a concise form for $\mu_\Phi$ (see {\bf Methods}), given by
\begin{align}
   \mu_{\Phi}= \frac{1}{N} + \left(\frac{1}{M}-\frac{1}{N}\right)
 \frac{1-e^{-\kappa\tau}}{\kappa\tau}. 
 \label{mu_phi_global}
\end{align}
We denote the CV value at $N=N^*$ by ${\rm CV}^*$.
By equating the first and second terms on the
right-hand side in \EQ\eqref{mu_phi_global} and 
assuming $M \ll N$ and $e^{-\kappa \tau} \ll 1$, we obtain
\begin{equation}
 N^* \approx \kappa \tau M , \quad {\rm CV}^* \propto \frac{1}{\sqrt{\kappa
  \tau M}}.
\end{equation}
Figure \ref{fig:kuramoto_global} shows the analytical and numerical results.
Note that in \FIGS\ref{fig:fhn_multi} and \ref{fig:hetero}(b),
the lower bounds
are roughly proportional to $1/\sqrt{M}$, as our theory predicts.

\begin{figure}[!ht]
\begin{center}
 \includegraphics[width=3in]{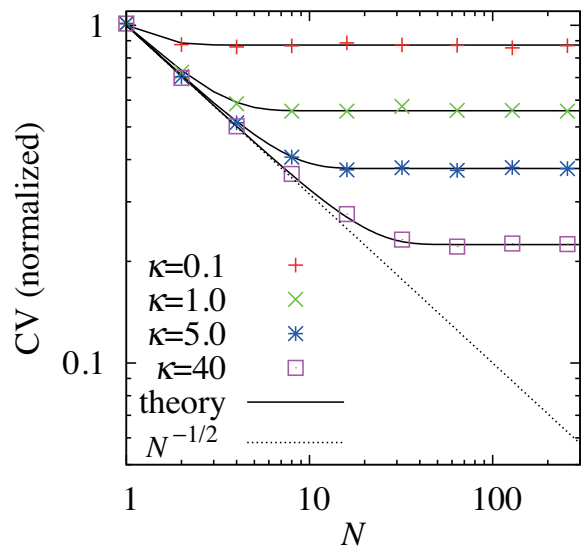}
\end{center}
 \caption{{\bf Normalized CV for single cells
in phase oscillators on the ring.}
 Symbols represent numerical results.
 Solid lines represent $\sqrt{\mu_i}$ given by \EQ\eqref{undirected2} with
 \EQ\eqref{eigen_ring}.
 }
 \label{fig:kuramoto_ring}
\end{figure}

{\em Ring.}
The third example is the ring of size $N$, 
i.e., 
$w_{i,i+1}=w_{i,i-1}=1/2$ for $1 \le i \le N$ and $w_{i,j}=0$
for $j\neq i-1, i+1$, as an example of spatially extended systems.
For this network, we obtain
\begin{equation}
 \lambda_n = 1-\cos\left(\frac{2(n-1)\pi}{N}\right) \label{eigen_ring}
\end{equation}
for $1 \le n \le N$.
%
%
Because $L$ is symmetric and the network has permutation symmetry, 
we obtain $\mu_{i}= \langle \mu \rangle$ where $\langle \mu \rangle$ is
given by \EQS\eqref{undirected2} and \eqref{eigen_ring}.  Figure \ref{fig:kuramoto_ring} shows
the analytical and numerical results. Although each cell is
adjacent to just two cells for any $N \ge 3$, there is a
clear $N$-dependence of the CV for individual cells.
Temporal precision is not simply determined by local connectivity.

\begin{figure}
\begin{center}
 \includegraphics[width=4in]{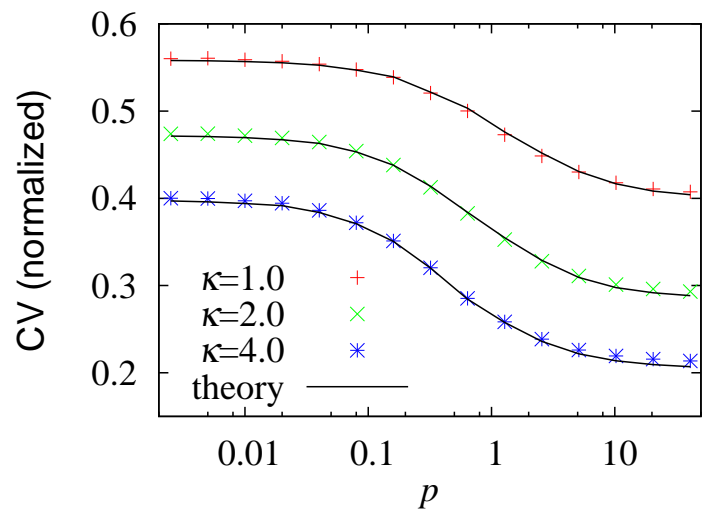}
\end{center}
 \caption{{\bf Normalized CV for individual cells
 in phase oscillators on the variant of the Watts-Strogatz model.}
 We present $\langle {\rm CV}_i \rangle / (\sqrt{\tau D}/2\pi)$, where
 $\langle {\rm CV}_i \rangle \equiv \sum_{i=1}^N {\rm CV}_i/N$.
 Lines represent $\sqrt{\mu_{i}}$ given by \EQ\eqref{undirected}.
We set $N=400$.}
 \label{fig:sw}
\end{figure}

%
The lower bound of the CV
for the ring is considerably
larger than that for the all-to-all network
(\FIGS\ref{fig:kuramoto_global}(a) and \ref{fig:kuramoto_ring}).
The reason for this is as follows.
The Laplacian of the ring for a large $N$ value has
negligible eigenvalues (i.e., $\lambda_n$
for $n\approx 0$ and $n \approx N$ in \EQ\eqref{eigen_ring}), 
and these eigenvalues significantly enlarge the second term of \EQ\eqref{undirected2}.
In contrast, 
there is a nonvanishing spectrum gap (i.e., the second smallest eigenvalue $\lambda_2$)
in the all-to-all and various random networks
\cite{monasson99,samukhin08}.
In the FHN model, we observed
a similar difference between the cases of the square lattice
(\FIG\ref{fig:fhn}(b)) and the all-to-all and random networks
(\FIGS\ref{fig:fhn}(a) and (c), respectively).
This is also because the square lattice has negligible eigenvalues \cite{mohar91}.
Such small eigenvalues are associated with
slow synchronization of remote oscillators owing to a time
lag in communication, and this property is shared by any
spatially extended networks with local interaction.
Therefore, spatial networks with only local interaction are
disadvantageous to temporal precision.
{\em Small-world networks.}
%
%
%
By using a type of the Watts-Strogatz model
\cite{newman00jsp,newman00prl} of fixed size $N$,
we demonstrate that a small fraction of long-range interactions
added to the ring drastically improves temporal precision.
We generate a network
by adding $pN$ bidirectional shortcuts sequentially
to the ring, where $p$ is the shortcut density. 
Under the condition that multiple links are avoided,
the two endpoints of each shortcut are chosen
from the $N$ nodes with equal probability.
The generated network is undirected.
%
%
To maintain the total coupling strength independent of $p$, we
set $w_{ij}=w_{ji}=1/(2+2p)$ for all links.
The ring and all-to-all networks
are obtained at
$p=0$ and $p=N/2-1$, respectively. Figure \ref{fig:sw}
shows the numerically obtained $\langle {\rm CV} \rangle$ for each $p$,
%
%
where $\langle {\rm CV} \rangle\equiv \sum_{i=1}^N {\rm CV}_i/N$
for a single realization of the network.
The lines represent
$\sqrt{\langle \mu \rangle}$ obtained from
\EQ\eqref{undirected2}, where we numerically computed the eigenvalues $\lambda_n$
for the generated network
We set
the coupling strength such that
$\langle \mu \rangle \gg 1/N$ (i.e., $N \gg N^*$)
for the initial ring ($p=0$). 

Figure~\ref{fig:sw} indicates that
temporal precision is considerably
improved at $p \approx 1$, i.e., when $O(N)$ shortcuts are added (the
small-world regime).
Moreover, the corresponding CV value is 
close to that of the all-to-all network,
in which $O(N^2)$ ``shortcuts'' exist.
As discussed above, 
there are small eigenvalues that hamper temporal precision in spatially
extended networks. 
Such small eigenvalues do not exist in networks with a sufficient number
of shortcuts because of rapid communication between any pair of oscillators.


\subsubsection*{Mechanism of the crossover}
We demonstrated using various models that the crossover generally occurs in the collective
enhancement.
On the basis of our theory, the crossover can be interpreted as follows.
When Jacobian $L$ is symmetric, the SD for the mean phase 
decreases as $1/\sqrt{N}$ for any $N$ (\EQ\eqref{mu_Phi_N}).
When coupling strength $\kappa$ is infinite, the oscillators are completely
synchronized in phase. Then, the phase of each oscillator is identical with
the mean phase, and so is
the SD, i.e.,
${\rm SD}_i \propto 1/\sqrt{N}$ for any $N$. 
This behavior is expressed in the first term on the right-hand side
of \EQ\eqref{undirected}.
However, for finite $\kappa$, individual oscillators'
phases fluctuate around the mean phase
because of the independent noise applied to
the oscillators. 
Owing to this additional fluctuation, the SD for individual
oscillators is larger than that for the mean phase, as expressed
in the second term on the right-hand side of \EQ\eqref{undirected}.
Although the fluctuation in the mean phase vanishes
with the limit $N\to \infty$,
it remains finite in individual oscillators.
This is the origin of the lower bound.

\section*{Discussion}
We found that 
the collective enhancement is ineffective for system size $N$ 
above the crossover point $N^*$.
We further showed that $N^*$ increases with coupling
strength (\EQ\eqref{N_propto_kappa}).
Therefore, as oscillators are more strongly coupled,
the behavior ${\rm CV} \propto 1/\sqrt{N}$ persists up to a larger $N$ value.
This is the case for different oscillation and
coupling mechanisms, as demonstrated in the 
two biological models (the FHN and SCN models) and the
phase oscillator model.
Moreover, this behavior also holds true for different network
connectivities, as demonstrated using the ring, the square lattice,
the all-to-all network, and the random graph.

Our theory is useful for inferring the magnitude of fluctuations in
individual cells and the coupling strength between cells.
Suppose that temporal precision in a pacemaker tissue
that is genetically modified or subjected to a treatment (e.g., drug)
is lower than that in an intact tissue.
If the cells in the tissue are well synchronized in both cases,
one may consider that the treatment affects the
oscillation mechanism of individual cells.
Our theory suggests another possibility:
a decrease in the coupling strength, not the alteration in the
oscillation property of individual cells, may be the reason for the
reduced temporal precision
(\FIGS\ref{fig:fhn}(a,b,c)).
By observing reduced temporal precision only,
we cannot distinguish these two
possibilities. 
However, our theory makes it possible to individually quantify
the effects of the treatment on the two properties 
if we can observe cell networks of
different sizes.
By observing temporal precision in small (i.e., $N<N^*$)
tissues 
of different sizes, we
can infer 
the magnitude of fluctuations in individual cells
by fitting the law ${\rm CV} \propto 1/\sqrt{N}$.
Furthermore, by observing relatively large tissues and determining
$N^*$ values for different treatments
(e.g., different days of cultivation,
different concentrations of a drug, treated versus untreated),
we can infer changes in the coupling strength induced by the treatment
because $N^*$ increases with the coupling strength (\EQ\eqref{N_propto_kappa}).


Our study also indicates that long-range interactions among cells
are advantageous to temporal precision.
As demonstrated in \FIG\ref{fig:sw}, the addition of shortcut links
considerably decreases the CV.
A similar result was reported in a previous numerical study using a more
realistic model for the SCN \cite{vasalou09}.
This result might underlie an evolutionary origin of
dense fibers across the SCN \cite{abrahamson01}.

Our theoretical results provide an interpretation
of previous experiments on cardiac and circadian oscillations.
Kojima {\em et al} observed 
a decrease in the CV with increasing cell number in cultivated
cardiac cells coupled via micro channels \cite{kojima06}.
They showed that the CV decreases considerably with $N$ for small $N$
values ($N=1,2,3$), 
while it is almost constant for $N \ge 4$.
In contrast, in cultivated
cardiac cells that are directly and tightly coupled to each other, 
Clay and DeHaan found that the reduction in the CV roughly obeys
${\rm CV} \propto 1/\sqrt{N}$ up to $N\approx 100$ \cite{clay79}.
Although the cells are kept synchronized in both cases, the behavior of
temporal precision is different.
This discrepancy may be due
to a difference in coupling
strength. While the coupling was strong enough to guarantee synchrony
in both cases,
coupling in the latter experiments 
may be stronger than that in the former experiments, resulting in
$N^* \approx 4$ and $N^* > 100$, respectively.
It would be of great interest to investigate systematically how the
crossover point increases with coupling strength, possibly controlled by
the width of the micro channel implemented in the former
experiments \cite{kojima06}.



Collective enhancement has been examined experimentally in
circadian oscillation as well.
Herzog {\em et al} measured temporal precision in SCN cells \cite{herzog04}.
There, individual cell oscillations in
both synchronized and unsynchronized cases were observed
in slice cultures of SCN and dispersed SCN cells, respectively.
They found that the SD in the former (0.42 h) was approximately five times
smaller than that in the latter (2.1 h),
and argued that the collective enhancement
of temporal precision occurs in synchronized cells.
They further speculated that, under the assumption ${\rm
SD} \approx 1/\sqrt{N}$, only 25 cells out of the order of $10^5$ cells
composing the SCN are involved in the collective enhancement of temporal
precision in the explant SCN. 

We interpret this experimental result as follows.
In the SCN, a wave pattern is observed \cite{yamaguchi03, doi11}.
As indicated above as well as in our
previous paper \cite{mkk10}, 
the law ${\rm SD}\propto 1/\sqrt{N}$
is violated 
in the presence of a wave pattern 
even if the coupling is sufficiently strong.
Roughly speaking, the reason for this is that only the cells forming
the source of the wave pattern can contribute
to the collective enhancement of temporal precision, and
other cells simply obey those cells \cite{mkk10}.
The number of cells forming the source might be of the order of 25.
Cells located downstream of the wave
may contribute to functions other than temporal precision.


Our theory is widely applicable to frequency-synchronized
oscillators with weak noise and weak coupling.
Our theory can also apply to the case of 
the coexistence of
multiple coupling mechanisms, 
only by replacing coupling function $f$ by $f_{ij}$ in the phase model (\EQ\eqref{model}).
Although the phase model is not justified
when the assumption of weak noise and weak coupling is violated, 
we have numerically confirmed that our main finding, i.e., the properties (i)--(v),
are preserved in the case of strong coupling and strong noise (\FIG\ref{fig:hetero}).
We thus expect that our theory, based on the phase model,
captures the essence of the collective enhancement of temporal
precision.
\section*{Methods}

\subsection*{Model equations for biological pacemaker systems}
We consider two systems---the FHN model and the SCN
model representing the cardiac pacemaker organ and the circadian master
clock, respectively.

The FHN model 
has been extensively used as a model
of neurons and cardiac cells \cite{keener98}.
Our FHN model is given by
\begin{subequations} 
\label{FHN}
 \begin{align}
 \frac{dx_i}{dt}&=x_i(a-x_i)(x_i-1)-y_i + \rho \xi_i(t) + \kappa \sum_{j=1}^N
  A_{ij} (x_j- x_i), \\
 \frac{dy_i}{dt}&= \epsilon (x_i- b y_i + c),
\end{align}
\end{subequations}
where $a,b,c,\epsilon$ are the model parameters, $\rho$ is the noise strength,
$\xi_i(t)$ is white Gaussian noise with 
${\rm E}[\xi_i(t)]=0$ and ${\rm E}[\xi_i(t)\xi_j(t')]=\delta_{ij}
\delta(t-t')$.
We chose parameter values such that each unit is autonomous
oscillator. In \FIGS\ref{fig:waveforms}, \ref{fig:fhn} and
\ref{fig:fhn_multi}, we set $a=0.1,\epsilon=0.01,b=0.5$, and $c=0.05$.
In \FIG\ref{fig:hetero}(a), we replace $c$ with
$c_i=0.1 + 0.02 \nu_i$ ($1\le i\le N$), where $\nu_i$ is a random
variable independently
taken from the Gaussian distribution with zero mean and unit variance.
We varied
the noise strength and coupling strength, as specified in the figures
and their captions.
The distance $\delta$ from the in-phase state is defined as
\begin{equation}
  \delta= \sqrt{ \frac{1}{N-1}\sum_{i = 1}^N (x_i - \overline{x} )^2},
\end{equation} 
where $\overline{x}= \sum_{i=1}^N x_i/N$.

As the SCN model, we employed a previously proposed
model \cite{locke08}, given by
\begin{subequations}
 \label{SCN}
\begin{align}
 \frac{dx_i}{dt}&= T_i \left(V_1\frac{K_1^n}{K_1^n+z_i^n} - V_2\frac{x_i}{K_2+x_i}
 + V_c\frac{\kappa F_i}{K_c+ \kappa F_i} \right) + \rho\xi_i^{(x)}\comb{,}\\
 \frac{dy_i}{dt}&= T_i \left( k_3 x_i - V_4\frac{y_i}{K_4+y_i} \right)+
 \rho\xi_i^{(y)}\comb{,} \\
 \frac{dz_i}{dt}&= T_i \left( k_5 y_i - V_6\frac{z_i}{K_6+z_i} \right)+
 \rho\xi_i^{(z)}\comb{,} \\
 \frac{dr_i}{dt}&= T_i \left( k_7 x_i - V_8\frac{r_i}{K_8+r_i} \right)+
 \rho\xi_i^{(r)}\comb{,} \\
 F_i &= \sum_{j=1}^N A_{ij} r_j,
\end{align}
\end{subequations}
where $V_1=6.8355,n=5.6645,K_1=2.7266,K_2=0.2910,k_3=0.1177,V_4=1.0841,K_4=8.1343,k_5=0.3352,V_6=4.6645,K_6=9.9849,k_7=0.2282,V_8=3.5216,K_8=7.4519,V_c=6.7924,K_c=4.8283,
\kappa=12.0$, and $V_2=12.0$.  
All parameter values except for $\kappa$ and $V_2$ are taken from \cite{locke08}.
Time constant $T_i$ is introduced to express heterogeneity in
the oscillation period. We set $T_i = 1$ in
\FIG\ref{fig:waveforms}. In \FIG\ref{fig:hetero}(b),
$T_i = 1 + 0.05 \nu_i$ with $\nu_i$
independently obeying
the Gaussian distribution with zero
mean and unit variance. 
The functions $\xi_i^{(\zeta)}(t)$ ($\zeta=x,y,z,r$)
represent white Gaussian noise processes with ${\rm
E}[\xi_i^{(\zeta)}(t)]=0$ and ${\rm E}[\xi_i^{(\zeta)}(t)
\xi_{j}^{(\eta)}(t')]=\delta_{ij} \delta_{\zeta \eta} \delta(t-t')$.
The noise strength $\rho$ and coupling strength $\kappa$ are specified in the 
figures and their captions.

In both models, we applied sufficiently strong
coupling to ensure that the oscillators were synchronized nearly in
phase.
When we computed the CV, we assumed random initial conditions and
measured a sufficiently large number of
cycle-to-cycle periods after the transient.

\subsection*{Networks}
\begin{figure}[!ht]
\begin{center}
\includegraphics[width=6.0in]{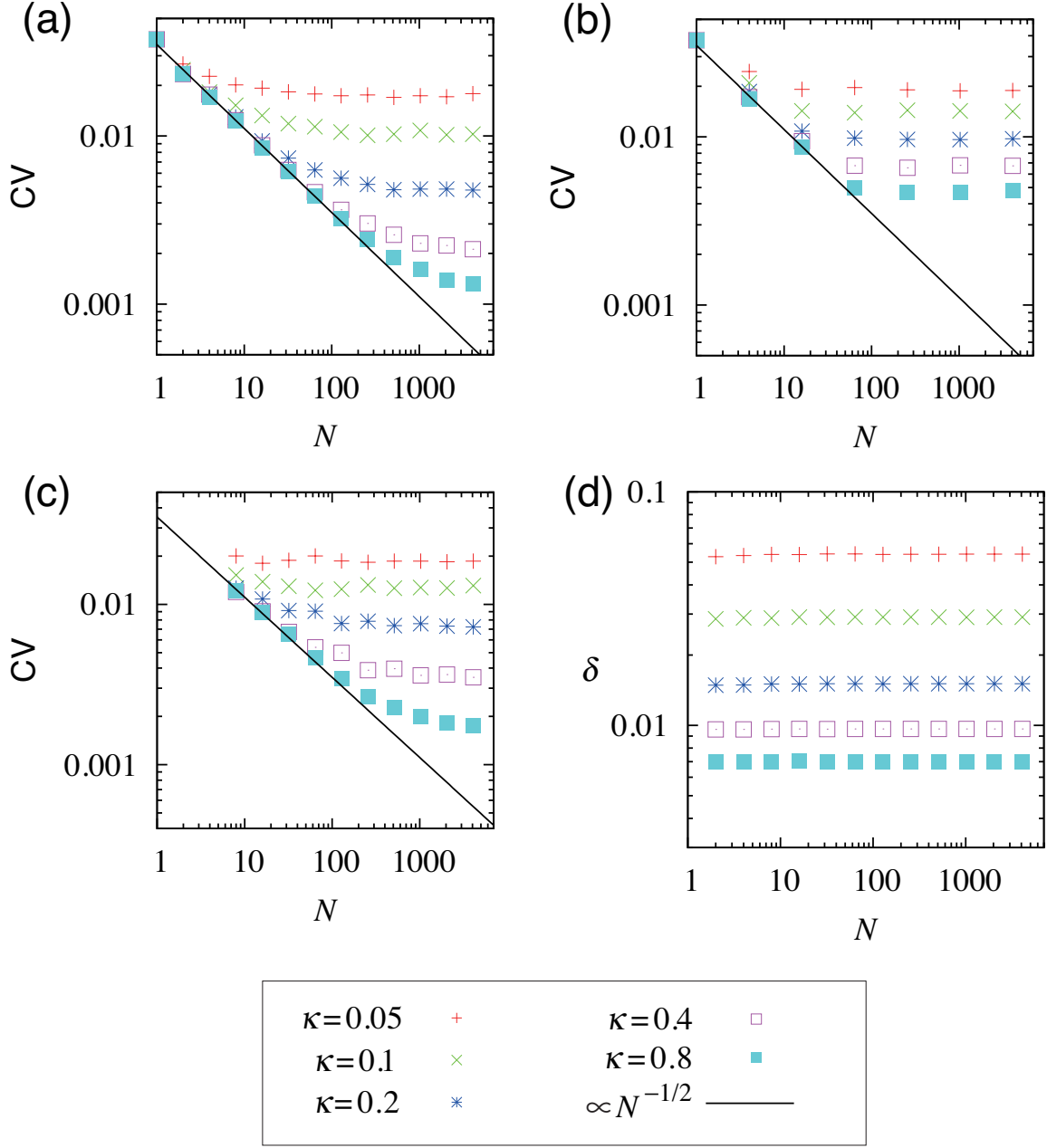}
\end{center}
\caption{
{\bf CV for single cell oscillations
and synchronization distance $\delta$ in the FHN model.}
(a,b,c) CV values for single cell oscillations on
(a) the all-to-all network, (b) the square lattice,
and (c) the undirected random graph
of size $N$.
(d) Distance $\delta$ from in-phase synchrony in the all-to-all
 network. In (a) and (d), we set $A_{ij}=1/N$ for $1\le i,j \le N$.
In (b), the CV of the oscillator 
at the center of the square lattice with an open boundary
condition is presented. We set
$A_{ij}=1/4$ with cell $j$ adjacent to cell $i$ and $A_{ij}=0$ otherwise.
In (c), the CV value at given $\kappa$ and $N$ values is defined as
$\langle {\rm CV}_i \rangle \equiv \sum_{i=1}^N {\rm CV}_i/N$
for a single realization of the network. We set
$A_{ij} = A_{ji} = 1/8$ with probability $p=8/N$ ($1\le i < j
 \le N$), and $A_{ij} = 0$ otherwise.
The lines are guides to the
 eyes.
We considered identical cells and weak noise ($\rho=0.01$).
The average period $\tau$ is almost constant $(\tau
\approx 177)$ irrespective of $N$ and $\kappa$.
}
\label{fig:fhn}
\end{figure}

\begin{figure}[!ht]
\begin{center}
\includegraphics[width=5in]{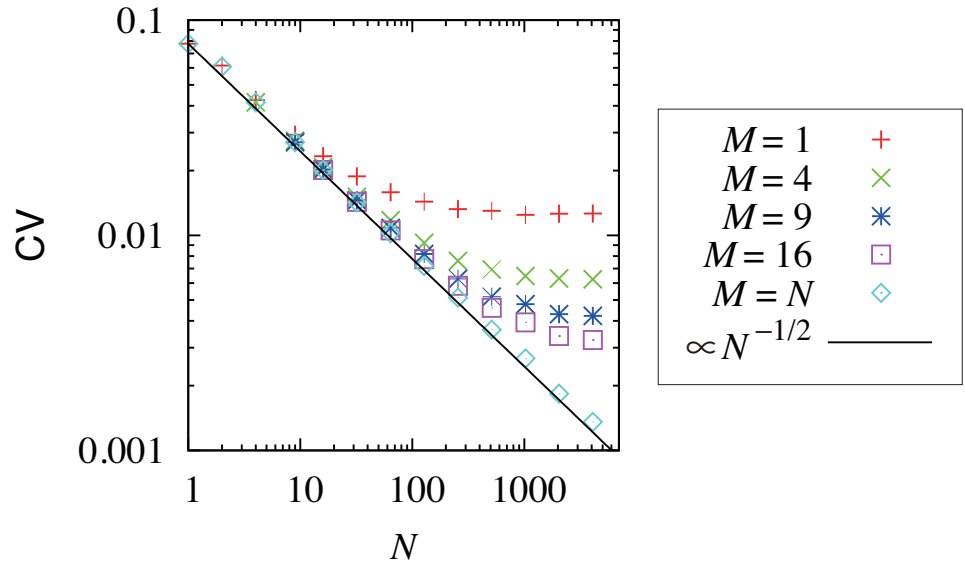}
\end{center}
\caption{
{\bf CV for ensemble activity of $M$ cells
in the FHN model on the all-to-all network.}
Parameter values are the same as in \FIG\ref{fig:fhn}, except $\rho = 0.0256$ and $K=0.2$.
The line is a guide to the eyes.
}
\label{fig:fhn_multi}
\end{figure}

\begin{figure}[!ht]
\begin{center}
\includegraphics[width=6in]{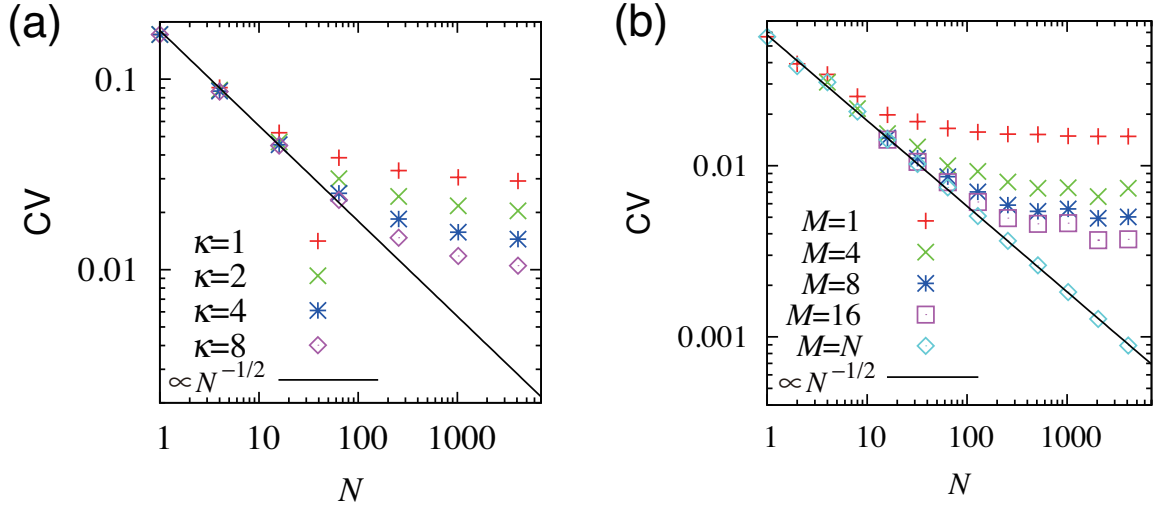}
\end{center}
\caption{
{\bf CV for biological models
composed of heterogeneous cells subjected to relatively strong noise.}
 (a) CV for single cell oscillations in the 
 FHN model on the square lattice. We set $\rho=0.09$.
 (b) CV for the ensemble activity
 of $M$ cells in the SCN model on the all-to-all network. We set $\rho=0.04$
 and $K=12$. 
The all-to-all network and the square lattice are the same as those in
\FIGS\ref{fig:fhn}(a) and (b), respectively.
The lines are guides to the eyes.
}
\label{fig:hetero}
\end{figure}

%
The all-to-all network used in \FIGS\ref{fig:waveforms}(b,d,f), \ref{fig:fhn}(a,d),
\ref{fig:fhn_multi}, \ref{fig:hetero}(b), and \ref{fig:kuramoto_global}
is defined by $A_{ij} = 1/N$ for $1\le i,j \le N$.
The one-dimensional lattice with an open boundary condition
used in \FIG\ref{fig:waveforms}(e) is defined
by $A_{ij}=1/2$ for $1\le i \le N$ and $1\le j=i \pm 1  \le N$, and $A_{ij}=0$ otherwise.
The ring used in \FIG\ref{fig:kuramoto_ring} is
the same as the one-dimensional lattice except that we impose a periodic boundary
condition $A_{1,N}=A_{N, 1}=1/2$.  The square lattice
with an open boundary condition used in \FIGS\ref{fig:fhn}(b) and
\ref{fig:hetero}(a) is defined
by $A_{ij}=1/4$ with cell $j$ adjacent to $i$
for $1\le i,j \le \sqrt{N}$ and
$A_{ij}=0$ otherwise.
The undirected random graph
used in \FIG\ref{fig:fhn}(c)
is the Erd\H{o}s-R{\'e}nyi
random graph,
where $A_{ij} =A_{ji}=1/8$ for $1\le i < j \le N$
with probability $p=8/N$ and $A_{ij} =A_{ji}=0$ otherwise. 
We set link weights such that the summed weight of
the links per node is independent of $N$; i.e.,
for $1 \le i \le N$, $\sum_{j=1}^N A_{ij} = 1$ in the ring and all-to-all network
and $\sum_{j=1}^N A_{ij} \approx 1$ in the other networks including the
Watts-Strogataz model used in \FIG\ref{fig:sw}.

\subsection*{Phase description}
A large class of oscillator systems including the 
FHN and SCN models (\EQS\eqref{FHN} and \eqref{SCN})
are reduced to phase models
if the coupling and noise are sufficiently weak \cite{winfree67,
kuramoto84}.
The concept behind the reduction is as follows.
We denote an element of the state
variable of the $i$th oscillator by $x_i(t)$. 
When unperturbed, 
the oscillator portrays
a one-dimensional closed orbit after transient so that
$x_i(t)=x_i(t+2\pi/\omega_i)$, where $\omega_i$
is the intrinsic frequency. We define the phase $\phi_i$
by $x_i(t)=x_i(\phi_i/\omega_i)$; that is, the phase increases linearly
with time in the unperturbed oscillator.
For convenience, we denote the unperturbed orbit by $\chi(\phi_i) =
x_i(\phi_i/\omega_i)$. 
Although the trajectory deviates from the closed orbit when the oscillator
is weakly perturbed,
it is still possible to parameterize a trajectory of an oscillator
by only the phase 
and describe the dynamics of coupled oscillators in terms of the phases
only \cite{winfree67, kuramoto84}.
The resulting equation is given by \EQ\eqref{model}.
Because of the assumption of weak perturbation,
$x_i(t)$ is approximated by that of the unperturbed
orbit, i.e.,
\begin{equation}
 x_i(t) \approx \chi(\phi_i(t)).
  \label{waveform}
\end{equation}
Therefore, the first passage time problem for $x_i(t)$ is approximated by
that for $\phi_i\CB(t)$.

\subsection*{Calculation of \EQ\eqref{C_ij}}
Our linearized equation is given by \EQ\eqref{linear}, which is
reproduced as
\begin{equation}
  \dot{\theta}_i(t) = -\kappa \sum_{j=1}^N L_{ij} \theta_j + \sqrt{D}\xi_i(t),
   \label{linear3}
\end{equation}
where $\theta_i$ ($1\le i \le N$) is the deviation from the synchronized state, $\kappa>0$ is the coupling strength, $L=(L_{ij})$ is a
diagonalizable matrix,
and $\xi_i(t)$ is white Gaussian noise with
\begin{equation}
  {\rm E}\left[ \xi_i(t) \right] = 0, \quad
  {\rm E}\left[ \xi_i(t) \xi_j(s) \right] = \delta_{ij} \delta(t - s).
\label{eq:def stat xi}
\end{equation}
From the assumption of the stability of frequency synchronization, we have
\begin{equation}
  0 = \lambda_1 < {\rm Re\,} \lambda_2 \leq {\rm Re\,} \lambda_3 \leq
   \cdots \leq \lambda_N. 
\end{equation}
The right and left
eigenvectors of $L$ corresponding to $\lambda_n$ 
are denoted by $\bm u^{(n)}=(u_i^{(n)})$ and $\bm
v^{(n)} = (v_i^{(n)})$, respectively; i.e., 
\begin{subequations} 
 \begin{align}
  L\bm u^{(n)} &=\lambda_n \bm u^{(n)},\\
  \bm v^{(n)} L &= \lambda_n \bm v^{(n)}
 \end{align}
\end{subequations}
with the normalization and orthogonality conditions
\begin{equation}
 \bm v^{(m)} \bm u^{(n)} = \delta_{mn}.
\label{ortho_diag}
\end{equation}
Using these eigenvectors, we decompose $\theta_i(t)$ as
\begin{equation}
  \theta_i(t) = \sum_{m=1}^N \varphi_m(t) u_i^{(m)}, 
   \label{theta_decomp}
\end{equation}
where $\varphi_m(t)$ is given by
\begin{equation}
 \varphi_m(t) = \sum_{i=1}^N \theta_i(t) v_i^{(m)}.
  \label{varphi}
\end{equation}
By taking the time derivative of \EQ\eqref{varphi} and using
\EQS\eqref{linear3}, \eqref{ortho_diag} and \eqref{theta_decomp}, we obtain
\begin{equation}
  \dot{\varphi}_m(t) = -\kappa \lambda_m \varphi_m + \eta_m(t),
   \label{dot_varphi}
\end{equation}
where 
\begin{equation}
  \eta_m(t) = \sqrt{D}\sum_{i=1}^N v_i^{(m)} \xi_i(t).
\end{equation}
Equation~\eqref{eq:def stat xi} yields
$\left\langle \eta_m(t) \right\rangle = 0$. 
We also have
\begin{align}
  {\rm E} \left[ \eta_m(t) \eta_n(s) \right]
  &= {\rm E} \left[ D \sum_{i=1}^N v_i^{(m)} \xi_i(t)
  \sum_{j=1}^N v_j^{(n)} \xi_j(s) \right]
  \nonumber \\
  &= D \sum_{i,j=1}^N v_i^{(m)} v_j^{(n)}
  {\rm E} \left[  \xi_i(t) \xi_j(s) \right]
  \nonumber \\
  &= D \sum_{i,j=1}^N v_i^{(m)} v_j^{(n)}
  \delta_{ij} \delta(t - s)
  \nonumber \\
  &= D \left(\sum_{i=1}^N v_i^{(m)} v_i^{(n)} \right) \delta(t - s)
  \nonumber \\
 &= D V_{mn} \delta(t - s),
\end{align}
where
\begin{align}
 V_{mn} \equiv \sum_{i=1}^N v_i^{(m)} v_i^{(n)}.
\end{align}
%

Now we derive $C_{ij}$ given in \EQ\eqref{C_ij}. The definition of
$C_{ij}$ is 
\begin{equation}
  C_{ij}  \equiv \frac{1}{D\tau} {\rm E}[(\theta_i(t + \tau) -
   \theta_i(t))(\theta_j(t + \tau) - \theta_j(t)) ].
\label{eq:C_ij SI def}
\end{equation}
By substituting \EQ\eqref{theta_decomp} in \EQ\eqref{eq:C_ij SI def}, we obtain
\begin{align}
  C_{ij}
  = \frac{1}{D\tau} \sum_{m,n=1}^N u_i^{(m)} u_j^{(n)}
  {\rm E} \left[ (\varphi_m(t + \tau) - \varphi_m(t) )
  ( \varphi_n(t + \tau) - \varphi_n(t) ) \right].
\label{C_ij SI}
\end{align}
%
%
The solution to \EQ\eqref{dot_varphi} is formally written as
\begin{equation}
  \varphi_m(t) = e^{-\kappa \lambda_m t} \varphi_m(0)
  + \int_0^t ds\, e^{-\kappa \lambda_m (t - s)} \eta_m(s).
  \label{solution}
\end{equation}
Using \EQ\eqref{solution}, we obtain
\begin{equation}
  \varphi_1(t + \tau) - \varphi_1(t) = \int_t^{t + \tau} ds\, \eta_1(s),
\end{equation}
\begin{align}
  {\rm E} \left[ (\varphi_1(t + \tau) - \varphi_1(t) )
  ( \varphi_1(t + \tau) - \varphi_1(t) ) \right]
  &= \int_t^{t + \tau} ds_1 \int_t^{t + \tau} ds_2 \,
  {\rm E} \left[  \eta_1(s_1) \eta_1(s_2) \right]
  \nonumber \\
  &= D V_{11} \int_t^{t + \tau} ds_1 \int_t^{t + \tau} ds_2 \, \delta(s_1 - s_2)
  \nonumber \\
  &= D V_{11} \tau.
 \label{B_11}
\end{align}
To evaluate the terms on the right-hand side of \EQ\eqref{C_ij SI}
for $m + n > 2$,
we first calculate
\begin{align}
  B_{mn}(t_1, t_2)
  &\equiv \int_0^{t_1} ds_1 \int_0^{t_2} ds_2 \, e^{-\kappa \lambda_m (t_1 - s_1) -\kappa \lambda_n (t_2 - s_2)}
  {\rm E} \left[ \eta_m(s_1) \eta_n(s_2) \right]
  \nonumber \\
  &= \int_0^{t_1} ds_1 \int_0^{t_2} ds_2 \, e^{-\kappa \lambda_m (t_1 - s_1) -\kappa \lambda_n (t_2 - s_2)}
  D V_{mn} \delta(s_1 - s_2)
  \nonumber \\
  &= D V_{mn} \int_0^{\min(t_1, t_2)} ds \, e^{-\kappa \lambda_m (t_1 - s) -\kappa \lambda_n (t_2 - s)}
  \nonumber \\
  &\to
  D V_{mn} \left.
  \frac {e^{-\kappa \lambda_m (t_1 - s) -\kappa \lambda_n (t_2 -
 s)}}{\kappa (\lambda_m + \lambda_n)} \right|_{s=\min(t_1, t_2)}.
 \label{B_mn}
\end{align}
We consider the limit $t\to \infty$ in \EQ\eqref{B_mn}
because we are concerned with a
stationary process.
%
Using \EQ\eqref{B_mn}, we obtain
\begin{align}
  & {\rm E} \left[ (\varphi_m(t + \tau) - \varphi_m(t) )( \varphi_n(t +
 \tau) - \varphi_n(t) ) \right] \nonumber \\
  &= B_{mn}(t + \tau, t + \tau) - B_{mn}(t + \tau, t) - B_{mn}(t, t + \tau) + B_{mn}(t, t) \nonumber \\
  &\to D V_{mn} \frac{2 - e^{-\kappa \lambda_m \tau} - e^{-\kappa
 \lambda_n \tau}}{\kappa (\lambda_m + \lambda_n)} \quad (t \to \infty).
 \label{B_mn_result}
\end{align}
By combining \EQS\eqref{C_ij SI}, \eqref{B_11} and \eqref{B_mn_result}, we obtain
\begin{equation}
  C_{ij}
  = V_{11} u_i^{(1)} u_j^{(1)}
  +  \sum_{m,n\; (m+n > 2)}^N
\frac{2-e^{-\kappa \lambda_m \tau}-e^{-\kappa \lambda_n
   \tau}}{\kappa(\lambda_m+\lambda_n)\tau}
 V_{mn} u^{(m)}_i u^{(n)}_j.
\end{equation}
In \EQ\eqref{C_ij}, we set $u_i^{(1)}=1/\sqrt{N}$ for $1\le i \le N$.

\subsection*{Scaling factor $\mu_\Phi$ for the ensemble activity}
In this section, we derive $\mu_\Phi$ used in
\EQS\eqref{mu_Phi_N} and \eqref{undirected_ensemble2}.
By substituting \EQ\eqref{waveform} in \EQ\eqref{ensemble_activity},
we express the ensemble activity $X(t)$ in terms of the phases as
\begin{equation}
 X(t)  = \sum_{i=1}^M \zeta_i x_i(t) \approx \sum_{i=1}^M \zeta_i \chi(\phi_i(t)).
\end{equation}
For in-phase synchrony (i.e., $\psi_i=0$) and small deviation
$\theta_i$, we can further approximate $X(t)$ to
\begin{equation}
 X(t) \approx \chi(\Omega t) + \sum_{i=1}^M \zeta_i \chi'(\Omega t)\theta_i(t)
  \approx \chi (\Phi(t) ),
\end{equation}
where $\chi'(\phi) = d \chi (\phi)/d\phi$ and $\Phi$ is the mean phase
of the ensemble, given by
\begin{equation}
 \Phi (t) = \Omega t+ \sum_{i=1}^M \zeta_i \theta_i(t).
\end{equation}
Thus, similar to the case of individual cell oscillations,
the cycle-to-cycle period for the ensemble activity $X(t)$
is approximated by the cycle-to-cycle period 
$\Delta t_\Phi^{(k)}$ for the mean phase $\Phi(t)$.
We further employ the following approximation (\FIG\ref{fig:heuristic})
\begin{equation}
 \frac{2\pi}{\tau}{\rm std}[\Delta t_\Phi] \approx {\rm std}[\Delta
  \Phi],
  \label{heuristic_Phi}
\end{equation}
where $\Delta \Phi \equiv \Phi(t+\tau) - \Phi(t) - 2\pi$.
%
We define the scaling factor $\mu_\Phi$ for the ensemble activity as
\begin{equation}
 {\rm var}[\Delta \Phi] = \mu_\Phi D \tau.
\end{equation}
We then obtain
\begin{equation}
 \mu_\Phi
 = \frac{{\rm var}[\Delta \Phi]}{D \tau} 
 = \sum_{i,j=1}^M \zeta_i \zeta_j \frac{{\rm E}[
(\theta_i(t+\tau)-\theta_i(t)) (\theta_j(t+\tau)-\theta_j(t)) ]}{D \tau}
 = \sum_{i,j=1}^M  \zeta_i \zeta_j C_{ij},
 \label{mu_Phi_general}
\end{equation}
where $C_{ij}$ is given by \EQ\eqref{C_ij}.

We consider the case of symmetric $L$ and $\zeta_i=1/M$ for
 $1 \le i \le M$.
Substituting \EQ\eqref{undirected} in \EQ\eqref{mu_Phi_general}, we obtain
\begin{align}
   \mu_{\Phi} = 
 \frac{1}{N} + \frac{1}{M^2} \sum_{n=2}^N \sum_{i,j=1}^M 
  \frac{1-e^{-\kappa \lambda_n \tau}}{\kappa \lambda_n \tau}
 u_i^{(n)} u_j^{(n)}.
 \label{undirected_ensemble}
\end{align}
For $M=N$, 
$\sum_{i=1}^{N} u_i^{(n)}=\bm u^{(1)} \cdot \bm u^{(n)}=0$ for $2\le n \le N$
(orthogonality) leads to
\begin{align}
   \mu_{\Phi} = \frac{1}{N},
\end{align}
that is, there is no crossover.
For $M < N$, \EQ\eqref{undirected_ensemble} implies that
$\mu_{\Phi}$ depends on the choice of $M$
oscillators. 
When we randomly choose $M$ out of $N$ oscillators, where $1\ll M \ll N$,
the dependence of $\mu_{\Phi}$ on $M$ is estimated as follows.
The orthogonality and normalization, respectively, imply
\begin{align}
  \frac{1}{N} \sum_{i=1}^N u_i^{(n)}=0, \qquad
  \frac{1}{N} \sum_{i=1}^N u_i^{(n)} u_i^{(n)}=\frac{1}{N}.
\end{align}
Therefore, the distribution of $u_i^{(n)}$ ($1\le i\le N$) 
has the mean of $0$ and variance of $1/N$.
We randomly choose $M$ ($\ll N$) elements and assume that they are
independent random numbers with the same mean and variance. 
Then, we apply the central limit theorem for $M \gg 1$ to obtain
\begin{equation}
    \sum_{i=1}^M \sum_{j=1}^M u_i^{(n)} u_j^{(n)}
    \approx \sum_{i=1}^M u_i^{(n)} u_i^{(n)}
    \approx \frac{M}{N}.
    \label{estimation}
\end{equation}
By substituting \EQ\eqref{estimation} in the right-hand side of
\EQ\eqref{undirected_ensemble}, we obtain \EQ\eqref{undirected_ensemble2}.


\subsection*{Calculation of \EQ\eqref{mu_phi_global}}
It is convenient to choose the eigenvectors $\bm u^{(n)}=(u_i^{(n)})$
for $2 \le n \le N$ as
$u_i^{(n)} = 1/\sqrt{n^2-n}$ for $1\le i \le n-1$, $u_n^{(n)} = (1-n)/\sqrt{n^2-n}$ and $u_i^{(n)} =
0$ for $n \le i \le N$.
Then, the following property holds:
\begin{equation}
 \frac{1}{M}\sum_{i=1}^M u_i^{(n)} = \left\{
 \begin{array}{cl} 
   0 & \quad \mbox{for $2\le n \le M$},\\
   \displaystyle{\frac{1}{\sqrt{n^2-n}} }  & \quad \mbox{for $ M+1 \le n \le N$}.
 \end{array} \right.
 \label{eigenvector}
\end{equation}
Substitution of \EQ\eqref{eigenvector} and
the eigenvalues $\lambda_n = 1$ ($2\le n
\le N$) in \EQ\eqref{undirected_ensemble} results in
\begin{align}
 \mu_\Phi
 &= \frac{1}{N} + \frac{1-e^{-\kappa \tau}}{\kappa\tau}
 \sum_{n=2}^{N} \left( \frac{1}{M} \sum_{i=1}^M u_i^{(n)} \right) \left( \frac{1}{M} \sum_{j=1}^M u_j^{(n)} \right) \nonumber \\ 
  &= \frac{1}{N} + \frac{1-e^{-\kappa \tau}}{\kappa\tau}
 \sum_{n=M+1}^{N} \frac{1}{n^2-n} \nonumber \\
 &= \frac{1}{N} + \frac{1-e^{-\kappa \tau}}{\kappa\tau}
 \sum_{n=M+1}^{N} \left( \frac{1}{n-1} - \frac{1}{n}\right) \nonumber \\
 &= \frac{1}{N} + \left( \frac{1}{M} - \frac{1}{N}\right)  \frac{1-e^{-\kappa \tau}}{\kappa\tau}.
\end{align}

\section*{Acknowledgments}
N.M. acknowledges the
support provided through Grants-in-Aid for Scientific Research (No. 23681033,
and Innovative Areas ``Systems Molecular Ethology'' (No. 20115009)) from MEXT, Japan.



\clearpage
\section*{Figure Legends}

\clearpage

\clearpage

\clearpage

\clearpage

\clearpage

\clearpage

\clearpage

\clearpage

\clearpage

\clearpage


\end{document}